\documentclass[12pt]{article}
\usepackage{epsfig}          % For eps figures, old commands
\usepackage[margin=2.5cm]{geometry}
\usepackage[utf8]{inputenc}
\usepackage{authblk}
\usepackage{graphicx}        % For eps figures, newer & more powerfull

\usepackage{color}           % For color text: \color command
\usepackage{breakurl}        % For breaking URLs easily trough lines
\chardef\us=`\_

\DeclareMathAlphabet{\mathsc}{OT1}{cmr}{m}{sc}
\def\testbx{bx}%
\DeclareRobustCommand{\ion}[2]{%
\relax\ifmmode
\ifx\testbx\f@series
{\mathbf{#1\,\mathsc{#2}}}\else
{\mathrm{#1\,\mathsc{#2}}}\fi
\else\textup{#1\,{\mdseries\textsc{#2}}}%
\fi}

\title{Future perspectives in solar hot plasma observations in the soft X-rays}
\author[a,*]{Alain Jody Corso}%\sep
\author[b]{Giulio Del Zanna}%\sep
\author[c,d]{Vanessa Polito}%\se

\affil[a]{\textit{\footnotesize{National Research Council of Italy - Institute for Photonics and Nanotechnologies, via Trasea 7, 35131 Padova (Italy)}}\vspace{1mm}}
\affil[b]{\textit{\footnotesize{DAMTP, CMS, University of Cambridge, Wilberforce Road, Cambridge CB3 0WA, United Kingdom}}\vspace{1mm}}
\affil[c]{\textit{\footnotesize{Bay Area Environmental Research Institute, NASA Research Park,  Moffett Field, CA 94035, USA}}\vspace{1mm}}
\affil[d]{\textit{\footnotesize{Lockheed Martin Solar and Astrophysics Laboratory, Building 252, 3251 Hanover Street, Palo Alto, CA 94304, USA}}\vspace{1mm}}
\affil[*]{\textit{\footnotesize{Corresponding author: alain.corso@pd.ifn.cnr.it}}\vspace{1mm}}

\date{}
\begin{document}

\maketitle
\hrulefill\par
\begin{abstract}
The soft X-rays (SXRs: 90--150~\AA) are among the most interesting spectral ranges to be investigated in the next generation of solar missions due to their unique capability of diagnosing phenomena involving hot plasma with temperatures up to 15~MK. 
Multilayer (ML) coatings are crucial for developing SXR  instrumentation, as so far they represent the only viable option for the development of high-efficiency mirrors in the this spectral range. However, the current standard MLs are characterized by a very narrow spectral band which is incompatible with the science requirements expected for a SXR spectrometer. Nevertheless,  recent advancement in the ML technology has made the development of non-periodic stacks repeatable and reliable, enabling the manufacturing of SXR mirrors with a valuable efficiency over a large range of wavelengths.  

In this work, after reviewing the state-of-the-art ML coatings for the SXR range, we investigate the possibility of using M-fold and aperiodic stacks for the development of multiband SXR spectrometers. After selecting a possible choice of key spectral lines, some trade-off studies for an eight-bands spectrometer are also presented and discussed, giving an evaluation of their feasibility and potential performance.    
\end{abstract}
\hrulefill\par


%-------------------------------------------------

\section{Introduction}
  The soft X-rays (SXRs: 90--150~\AA) have special diagnostic capabilities to
probe hot plasma, from about 3 to 15~MK (see, e.g. the \cite{delzanna_mason:2018} review).
In this range, the spectra not only contain strong lines from
six ionization stages of iron (from \ion{Fe}{xviii} to \ion{Fe}{xxiii}), but they also
have unique diagnostics to measure electron densities of such hot plasma. 
The SXRs have therefore an unique potential
to study a wide range of phenomena such as  microflares, nanoflares, and especially non-equilibrium ionization (NEI) at such temperatures. In particular, the study of the heating and cooling cycles with NEI requires the simultaneous observation of several ionization stages and, at the same time, accurate measurements of electron densities. The importance of the SXRs  was briefly mentioned in a white paper submitted in 2016 
by one of us (GDZ) in response to an international call to provide suggestions for a Next Generation JAXA/NASA/ESA solar physics mission.
It was suggested that the SXRs should be 
considered in future missions, as they are the best option to probe the cooling and heating of hot (5--10 MK) plasma in the solar corona, which is fundamental to try and resolve the coronal heating problem in active regions.
%The same view was shared by others, including J. Klimchuk, a leading expert on the problems related to  solar coronal heating  (cf. \cite{klimchuk:06}).

So far, existing spatially-resolved spectroscopic observations have been
mostly focusing on the extreme ultraviolet (EUV: 150--400 \AA) and the  ultraviolet (UV: 400 --2000 \AA) spectral regions.
Most  science missions to study the solar corona in the past 40 years have
focused on the EUV range, because of its excellent diagnostics, in particular in the lower-temperature plasma up to about 4 MK. We note that the EUV also includes some lines probing 
the very hot flare plasma at 15~MK (from \ion{Fe}{xxiii} and \ion{Fe}{xxiv}), but the lines at intermediate temperatures are weak or blended \cite{delzanna:08_bflare}.
%In the EUV, observations have been mostly with normal-incidence multilayers (ML), as e.g. in the case of the Hinode EIS instrument \cite{Culhane:2007},  mainly because of the availability of EUV multilayers which provide excellent reflectivities. Normal incidence mirrors can also be polished to achieve very good spatial resolutions. For example, an EUV multilayer at 171~\AA\ was employed by the Hi-C sounding rocket, achieving a spatial resolution of about 0.25 arcseconds \cite{podgorski_etal:2012_hic,kobayashi_etal:2014}.
In the UV range there are five strong hot lines between 592 and 1355~\AA: \ion{Fe}{xviii} (975~\AA), \ion{Fe}{xix} (592 and 1118~\AA), \ion{Fe}{xx} (722~\AA), and  \ion{Fe}{xxi} (1354~\AA). These lines have comparable intensities as those in the SXRs, but are far apart in wavelength (i.e. can be difficult to measure by a single instrument).
%, and lack of density diagnostics. 
Such hot lines have been observed by several instruments to study heating in active regions, such as the Solar Ultraviolet Measurements of emitted radiation (SUMER) \cite{parenti_etal:2017} on board \textit{SOHO} and, more recently, the \textit{Interface Region Imaging Spectrometer (IRIS)}  \cite{DePontieu14,polito_etal:2015}.
% The drawback of previous instruments in the UV has been the very low signal. 
 Some of these lines will also be observed with much higher throughput by the \textit{EUV Spectroscopic Telescope (EUVST)} \cite{Shimizu19}, an M-class mission proposed to the Japanese Space Agency (JAXA). The \textit{EUVST} design is based on the LEMUR instrument \cite{teriaca_etal:2012_lemur}: the optical components have a 
standard ML coating covering the 170-–215~\AA, and a B$_4$C top layer
providing good reflectances in three UV bands: 690–-850 \r{A},
925–-1085 \r{A}, and 1115–-1275~\AA. 

A valuable alternative to the EUV and UV is the X-rays (5--50~\AA) range. This spectral region is rich of lines emitted by hot plasma, although instrumentation working in this spectral range is based on grazing incidence optics which can add some challenges (such as smaller effective areas and moderate spatial resolution).  
%Also, previous missions have lacked any spatial resolution. A significant improvement is achieved with grazing incidence  focusing optics, but the spatial resolution obtained so far has been limited, being around  5 to 10 arcseconds.
%For instance, the FOXSI concept \cite{krucker_etal:2014} has flown on sounding rockets, achieving a large effective area, at the expenses of spatial and energy resolution. Further, the MaGIXS concept  \cite{kobayashi_etal:2011}, due to fly on a sounding rocket, will provide the first spatially-resolved X-ray spectra of the Sun, but at the expense of lower signal.

 The SXRs still remain largely unexplored, as no instrument has provided spatially-resolved spectroscopy in this range so far. However, we note that there has been renewed interest in SXRs among the solar physics community. For instance, the latest version of the \textit{Extreme Ultraviolet Normal Incidence Spectrograph (EUNIS--2)} will provide the first spatially-resolved SXR spectra of the Sun. \textit{EUNIS--2} is a two-channel imaging spectrograph (89-–112 \r{A} and 520-–640~\AA) to be launched on a sounding rocket in 2020.  Another example is given by the \textit{Multi-slit Solar Explorer (MUSE)} \cite{depontieu_etal:2020}, a
proposed MIDEX mission. \textit{MUSE} is based on a novel multi-slit design using 37 slits, that will allow to obtain simultaneous SXR/EUV spectra and images at the unprecedented spatial (0.33--0.4 arcseconds) and temporal (1--4 s) resolution for the transition region (TR) and corona.
The \textit{MUSE} design includes three passbands, with one channel observing the hot \ion{Fe}{xix} 108.36~\AA\ 
and the \ion{Fe}{xxi} 108.12~\AA\ lines. Finally, the next generation of the \textit{High-resolution Coronal Imager (Hi-C)} Rocket Experiment \textit{(Hi-C FLARE}) will include a primary telescope and a slitless spectrometer (COOL-AID) \cite{Winebarger19} observing the \ion{Fe}{xxi} line at 129~\AA. 

The development of future SXR instrumentation should take into account the expertise acquired in recent decades from previous EUV missions. In particular, EUV observations have mostly been performed with instruments based on a normal-incidence configuration, as for instance in the case of the EUV Imaging Spectrometer (EIS) \cite{Culhane:2007} on board \textit{Hinode}, thanks to the availability of EUV MLs that provide excellent reflectivities in this spectral range \cite{Corso:2019}. One of the biggest benefits of the normal incidence configuration, as compared to grazing incidence, is the possibility of achieving large effective areas. In fact, apart from a small central region that in on-axis schemes is obstructed by the secondary mirror, normal incidence instruments exploit the whole diameter of the primary mirror for collecting light. 
%In contrast, grazing incidence systems can only collect radiation in the small annular region subtended by the primary mirror \cite{barstow:2003}. 
Moreover, normal incidence mirrors can be polished to achieve very good spatial resolutions. For example, a EUV ML at 171~\AA\ was employed by the \textit{Hi-C} sounding rocket, achieving a spatial resolution of about 0.25 arcseconds \cite{podgorski_etal:2012_hic,kobayashi_etal:2014}. Luckily, the ML technology allows the realization of high-performing normal incidence mirrors also in the SXRs \cite{Corso:2019}. Besides, 
the great potential of using SXRs to study hot plasma has been recognised 
by the solar community thanks to two SXR  Y/Mo and Mo/Si MLs, at  94 \r{A} and 131~\AA,
used by the Atmospheric Imaging Assembly (AIA) \cite{Lemen:2012} on board the \textit{Solar Dynamic Observatory} since 2010. The AIA telescopes are providing 
high resolution ($\approx$~0.6~arcseconds) images with good signal to noise,
thanks to a  peak reflectance of 
0.4 and 0.7 \cite{soufli_etal:2005}, obtained at the expense of
a narrow spectral band.
Additionally, the telescope Zr filters have suffered very little degradation \cite{Boerner:2014}.

In summary, a SXR spectrometer working in normal-incidence would be highly desiderable.
So far the main limitation for the development of such a SXR spectrometer 
has been the very narrow spectral range achieved by 
standard MLs. Therefore,  it is crucial to explore
other options which can provide high peak reflectances over a large range of wavelengths.
This is the focus of the present paper, which is organized as follows: in Section 2, a general overview of the available and most promising materials for the SXR MLs is presented. In Section 3, we review the state-of-the-art of different types of ML stacks (e.g. M-fold, aperiodic structures) as well as their performance. Finally, in Section 4 we discuss a possible choice of key spectral lines, and some trade-off studies for the development of a possible multi-band spectrometer.

\section{SXR materials selection}
\begin{figure} [ht]
   \begin{center}
   \begin{tabular}{c} 
   \includegraphics[width=0.96\textwidth]{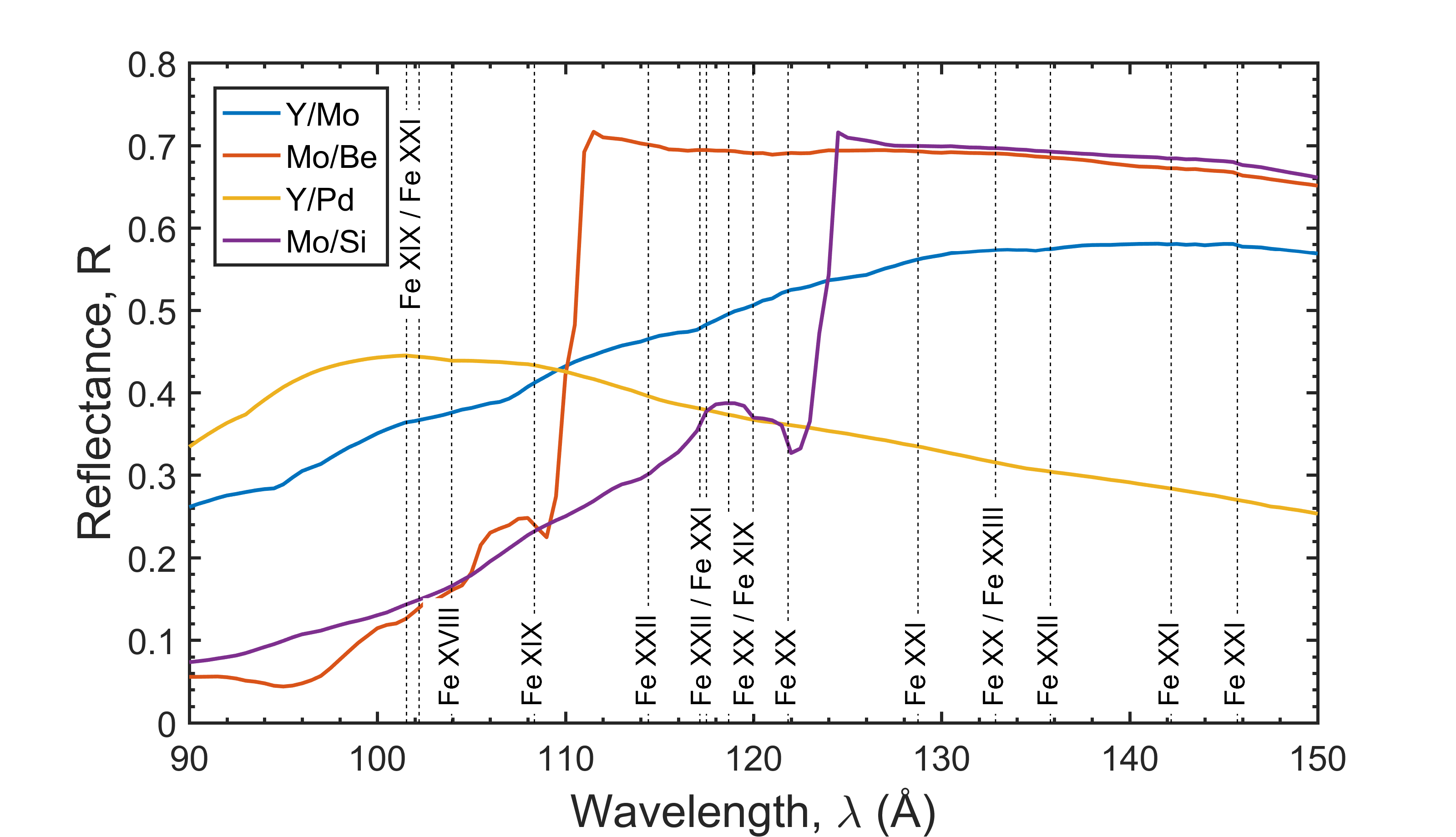}
	\end{tabular}
	\end{center}
   \caption[Summary of the reflectance performance for the most promising MLs in the 90 - 150 \r{A} wavelength range.] 
   { \label{fig:MLsummary} 
Simulations of the maximum reflectance (i.e. native efficiency) expected from Y/Mo, Mo/Be, Y/Pd and Mo/Si periodic stacks at 5$^\circ$ incidence angle in the 90 - 150 \r{A} wavelength range. The simulations were performed using the optical constants available in the CXRO database \cite{CXRO:web}. The effects of the interfaces roughness and inter-diffusion were modelled using a Gaussian error function according to the N\'evot and Croce formalism \cite{Nevot:1980}. The Y/Mo interfaces were modelled with an error function with $\sigma =0.57$ nm \cite{Corso:2015}. For Mo/Be, two different error functions were used: one with a $\sigma =0.36$ nm for Be on Mo interfaces and $\sigma =0.7$ nm for Mo on Be interfaces \cite{Svechnikov:2018, Bajt:1999SPIE}. Finally, for the Pd/Y and Mo/Si structures interface roughness error functions with $\sigma =0.85$ nm \cite{Windt:2015} and $\sigma =0.65$ nm \cite{Corso:2011} were used, respectively. }
\end{figure}

The development of highly performing MLs working in the SXR (i.e. 90 - 150 \r{A}) spectral range is particularly challenging. The thin stack period required for working at these wavelengths (i.e. roughly a quarter of wavelength) makes the selection of material couples having both good chemical/mechanical compatibility and high optical contrast quite difficult. So far only a few structures have shown high optical performance as well as high promising stability over time and thermal cycles \cite{Corso:2019}. In order to compare the performance of such structures, the native efficiency of the materials couple is typically used; this is defined as the maximum reflectance achievable in a given wavelength by an optimized periodic stack. The native efficiency of the most promising SXR MLs is reported in Figure \ref{fig:MLsummary}. 

For wavelengths longer than the Si L$_{2,3}$-edge ($\lambda \sim $124 \r{A}), the highest normal incidence reflectance is given by Mo/Si stacks. In fact, at these wavelengths the Si is almost fully transparent, and thus it acts as a quasi-perfect spacer material. In the SXRs, the reflectance which is attainable with such structures approaches 70\% for wavelengths just above the L$_{2,3}$-edge and remains above 65\% up to 160 \r{A} \cite{Corso:2019}; the spectral band of a periodic stack tuned in this range is, however, as narrow as $\frac{\Delta\lambda}{\lambda}<4\%$. Moreover, Mo/Si provides high stability and thermal resistance over time, making it very suitable for instruments on board long-term solar missions, such as the Extreme-ultraviolet Imaging Telescope (EIT) \cite{Delaboudiniere:1995}, on board \textit{SOHO}, \textit{Hinode}/EIS and \textit{SDO}/AIA.

Very recently, there has also been a renewed interest in beryllium-based MLs \cite{Svechnikov:2018}. In the past, such structures were investigated for the development of high-efficiency mirrors for the new generation EUV photo-lithography apparatus \cite{Bajt:2000}. However, their use has been progressively decreased  due to the serious safety issues arising when this material is handled (i.e. beryllium is highly toxic and carcinogenic). Nevertheless, Be is the only transparent material for wavelengths $>$111 \r{A} (Be K-edge) and is as transparent as Si around 135 \r{A} \cite{Svechnikov:2018}. Therefore, Be is a valuable spacer material for ML structures tuned at wavelengths close to its K-edge. The Mo/Be stack is a low-stress structure which shows the highest performance in the 111 \r{A}--124 \r{A} wavelength range: at 112.8 \r{A}, it provides a reflectance of 70.2\%  at an angle of 6$^\circ$ from the surface normal \cite{Svechnikov:2018}. The structure terminates with the top-most Be, with consequent formation of a native BeO layer of 3-4 nm. This seems to be the most stable structure over time: an absolute reflectance loss of 2\% has been observed after 20 months of storage at room temperature in air, whereas the termination with Mo or some other metallic capping-layer induces a greater reflectance loss \cite{Bajt:1999SPIE}. The only exception is given by the use of a Mo$_2$C layer with the addition of a very thin carbon layer (0.2 nm) on top of the stack. Thanks to this addition, the structure showed an absolute reflectance loss of 0.7\% after ten months \cite{Bajt:1999SPIE}. Moreover, Mo/Be seems to be stable also after fast thermal cycles in argon atmospheres, where 30 s of temperature rise are followed by 30 s of temperature maintenance and 3 minutes of cooling down time. No change in reflectance was detected for samples heated between 82 $^\circ$C and 239 $^\circ$C, while absolute reflectance losses of 0.7\% and 1.5\% were detected at 293 $^\circ$C and at 343 $^\circ$C, respectively \cite{Bajt:1999SPIE}.

At wavelengths shorter than the Si L$_{2,3}$-edge, a number of different material couples can be considered \cite{Montcalm:1996}. In particular, Y-based MLs have demonstrated good performance in the $\lambda \sim $80--120 \r{A} spectral range because of the proximity of the Y M$_{4,5}$ absorption edge at $\lambda \sim $ 79.5 \r{A}. However, among many proposed Y-based MLs, only a few appeared to be stable and suitable for space applications \cite{Corso:2019}. The most famous and studied one is the Y/Mo ML, since it shows very good aging and thermal stability, low stress \cite{Windt:2004, Kjornrattanawanich:2004, Xu:2015_ThinSolidFilms} and a relatively high reflectance peak (33\% - 35\% close to the \ion{Fe} {xviii} line at $\sim$94 \r{A} \cite{Sae-Lao:2002, Windt:2015}, 40\% at 105 \r{A} \cite{Windt:2015} and 46\% at 114 \r{A} \cite{Montcalm:1995}). Thanks to its good preformance, Mo/Y ML coatings have been used in long-term space solar observatories, including \textit{SDO}/AIA and the Solar Ultraviolet Imager (SUVI)\cite{Martinez-Galarce:2013} on board \textit{GOES-R}.  A very promising alternative ML in the 80--120 \r{A} wavelength range is provided by the Y/Pd stack \cite{Windt:2015, Xu:2015_OE}. The dramatic decrease of reflectance induced by the severe intermixing which occurs between the Pd and Y layers can in fact be overcome by adding a sub-nm-thick B$_4$C barrier layer at each interface of a Y/Pd ML \cite{Windt:2015}. Such structures provide high reflectance at 94 \r{A} (i.e. 43\% with an incidence angle of 5$^{\circ}$), good stability over time and thermal stability up to 100 $^{\circ}$C. So far, periodic Y/Pd ML is the coating that gives the highest performance in the 90-110\r{A} spectral range, although it is characterised by a very narrow spectral-band (i.e $\frac{\Delta\lambda}{\lambda}\simeq2 - 3\%$).

\section{Non-periodic SXR ML coatings: an updated overview}
Most of the the solar missions so far have employed standard periodic ML coatings for developing EUV/SXR mirrors with high reflectance at normal incidence. Such coatings have been studied for over 40 years and nowadays they represent a reliable technology in terms of performance, thicknesses precision, reproducibility and stability over time. However, periodic stacks are characterized by a reflectance spectral curve which shows a high and narrow peak at a specific target wavelength $\lambda$; thus, such MLs have narrow spectral bands, usually $\frac{\Delta\lambda}{\lambda}<10\%$ and even $\frac{\Delta\lambda}{\lambda}<5\%$ for wavelengths $< 150$ \r{A}, which is particularly limiting for some solar physics applications. For instance, spectroscopic observations usually require broadband coatings with high reflectance over the whole spectral band. This is even more important for spectroscopy in the SXRs, a spectral range that is still practically unexplored and hard to access, where the standard ML coatings give high reflectance in a very narrow spectral-band. Unfortunately, enlarging the spectral-band of a SXR ML always results in a drop of the maximum peak, which becomes dramatic when the reflectance curve is flattened. Therefore, it is often necessary to find a trade-off between the desired reflectance, spectral bandwidth and shape. In solar physics, an interesting compromise for SXR spectroscopy is the use of multiband mirrors tuned to the emission lines required for a specific plasma diagnostic. In this way, the simultaneous observation of different emission lines is possible without degrading too much the reflectance at these wavelengths. Such coatings can be made by using non-periodic MLs. 
%For spectroscopy, having a very narrow spectral range is a strong limiting factor, partially mitigated in the EUV where good reflectances are easily achieved over 2-3 nm ranges. 

In the last decade, a significant advancement in EUV/SXR non-periodic coatings has been led by the photolithography industry and the large scale facilities, such as Free Electron Lasers (FELs) and synchrotron beam-lines. Based on the success of non-periodic ML structures in these research fields, it seems appropriate to propose the use of these coatings for future solar missions. In the following sub-section, we present a review of non-periodic SXR MLs.

\subsection{M-fold multilayers}
In the last two decades, the two-fold EUV/SXR ML mirrors - a structure based on the superposition of two periodic stacks - have been mainly proposed either for broadening the native-spectral band of a periodic structure \cite{Corso:2015,Morlens:2006} or for developing multi-band mirrors \cite{Gautier:2008, Hecquet:2009}. For instance, overlapping two periodic structures whose period is slightly varied will produce a total response with high reflectance in a wider bandwidth, roughly given by the superimposition of the bandwidth of each stack \cite{Corso:2015}. Similarly, if the two periodic MLs are tuned at two wavelengths far from each other, a dual-band ML can be obtained \cite{Gautier:2008}. This strategy has been exploited for the Full Sun Imager (FSI) and the High-Resolution Imager (HRI) channels of the \textit{Extreme Ultraviolet Imager} (EUI) instrument onboard the ESA \textit{Solar Orbiter (SO)} mission \cite{Halain:2010, Rochus2020}. 
The HRI and FSI MLs have been optimized to obtain high reflectance at 174 \r{A} (\ion{Fe}{x}) / 335 \r{A} (\ion{Fe}{xvi}) and 174 \r{A} / 304 \r{A} (\ion{He}{ii}) lines, respectively \cite{Hecquet:2009}. The telescope optical path is shared by both wavelength channels and the selection of one or the other band is simply carried out by thin metallic filters (i.e. Zr for \ion{Fe}{x} and Al/Mg/Al for \ion{He}{ii} / \ion{Fe}{xvi} lines) \cite{Halain:2010}. 

The generalization of the two-fold ML is the M-fold ML, a structure composed by the superposition of M periodic blocks, each tuned at a different wavelength. In the EUV/SXR range, the number of blocks is usually limited to 3 or 4, mainly due to the material absorption. Moreover, in the most general case, the M-fold ML has also some additional “matching layers” between the periodic blocks and on top of the structure. Such layers, intended to match the different equivalent electromagnetic admittance given by the periodic sections, provide grater flexibility during the design process and therefore enhanced performance.

M-fold MLs are clearly very suitable for developing multi-band systems, where high reflectance peaks are required only for a discrete set of wavelengths and there is not a particular requirement for the  shape of the reflectance curve. Although the reflectance peak given at each target wavelength is lower than that achievable with other strategies (i.e. the aperiodic structures discussed in the following sub-section), M-fold MLs are robust and reliable in terms of deposition errors. In fact, the modern and sophisticated deposition facilities as well as the calibration strategies developed for the photo-lithography industry nowadays ensure very high precision in the manufacturing of periodic stacks \cite{Louis:2011}. Since M-fold MLs are based on overlapping periodic stacks, the final reproducibility and reliability achieved in the deposition of such structures is still very high, as demonstrated, for example, by the results obtained for \textit{SO}/EUI. Therefore the use of M-fold MLs can be considered a low-risk strategy for the next generation of solar missions. 

\subsection{Depth-graded and aperiodic multilayers}
\begin{figure} [!ht]
   \begin{center}
   \begin{tabular}{c} 
   \includegraphics[width=0.96\textwidth]{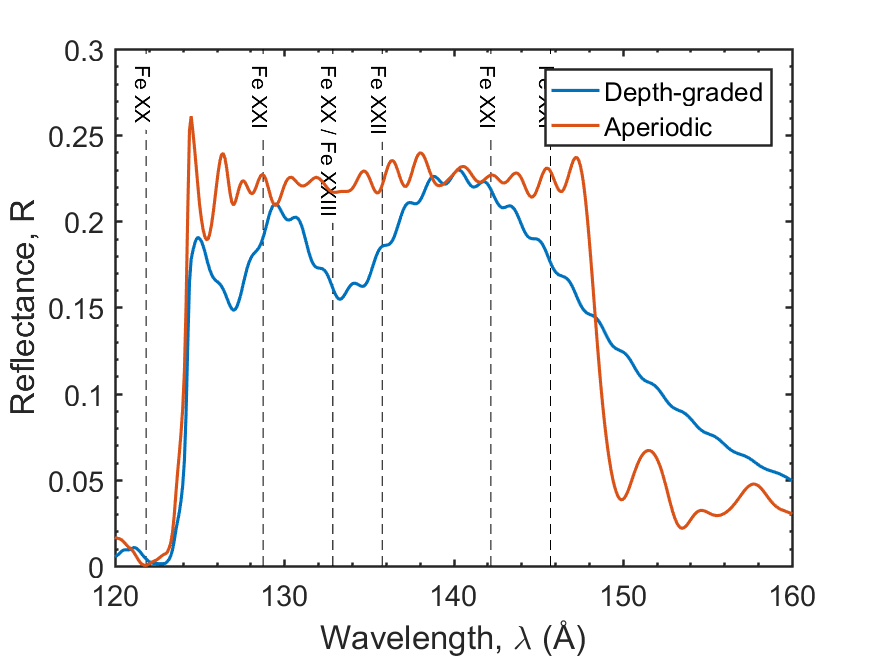}
	\end{tabular}
	\end{center}
   \caption[Comparison between the broadband reflectance shape given by a Mo/Si  depth-graded and that of a fully-aperiodic ML in the 124-148 \r{A} wavelength range.] 
   { \label{fig:MLlogape}  Comparison between the broadband response obtained with a logarithmic depth-graded Mo/Si ML (blue curve) and an aperiodic ML (red curve). Simulations at 5$^\circ$ of incidence were performed using the optical constants available in the CXRO database \cite{CXRO:web} and assuming an interface roughness of $\sigma=0.65$ nm.}
\end{figure}

In many applications, such as X-ray astronomy \cite{Yao:2013}, X-ray imaging \cite{Pardini:2016}, EUV photolithograpy \cite{Feigl:2006} and ultrashort pulses manipulation \cite{Wonisch:2006}, the need of mirrors with broad spectral or angular bandwidth has led to the development of depth-graded \cite{Lee:1983} and aperiodic \cite{Suman:2007} MLs. The term "depth-graded" denotes a structure where the period has been analytically varied along the depth of the stack following a monotone function. In this way, each bi-layer of the stack produces constructive interference for a slightly different wavelength, allowing to achieve high reflectance in a broad spectral range. This concept was firstly proposed for the design of X-ray \cite{Lee:1983} and neutron super-mirrors\cite{Mezei:1989} and, later, for EUV/SXR applications \cite{Vkozhevnikov:2001}. Many examples of depth-graded MLs are available in the literature, such as those reported in \cite{Windt:2015}, based on the Y/Mo and Y/Pd structures. In particular, structures having their period varied according to a logarithmic law are most promising for broadening the spectral band of the MLs, although the resulting shape of the band will not be flat. An example is reported in Figure \ref{fig:MLlogape}, showing a Mo/Si ML which has been optimized to obtain a broadband structure in the 128--148 \r{A} wavelength range. This ML structure has been graded using the following formulas: $x_{Si}=55-4.93\ln{2i}$ and $x_{Mo}=0.56x_{Si}$.

A substantial improvement, both in terms of maximum reflectance and flatness of the band shape, can be achieved by using fully aperiodic structures, in which the thickness of each layer is independent of each other. In fact, a larger number of degrees of freedom (i.e. the number of layers in the stack) provides higher flexibility during the design process, and thus allows to obtain reflectance curves with generic shapes \cite{ Windt:2015, Aquila:2006, Suman:2007}. An example is reported in Figure \ref{fig:MLlogape}, showing a Mo/Si logarithmic depth-graded  ML which has been made aperiodic to improve the flatness of the reflectance curve. 
The aperiodic approach can also be employed to improve the performance of M-fold structures, especially for the development of multiband mirrors. An interesting example has been reported by \cite{Gullikson:2015}, who used a fully aperiodic Mo/Si stack to reflect the FEL harmonics at 131.1 \r{A}, 135.6 \r{A} and  140.3 \r{A}, with an efficiency of more than 35\%.

However, depth-graded and aperiodic structures have some drawbacks that is worth mentioning. Firstly, the achievement of the desired EUV/SXR performance requires very high accuracy in the thickness of the layers, since even small errors can result in large changes to the ML reflectance curve. Thus, it is necessary to calibrate  accurately the individual deposition rates for each layer in order to minimize any errors associated with the deposition process. As a consequence, the manufacturing of these structures is not very easily reproducible. A second important aspect concerns the optimization procedure. The design of fully aperiodic MLs is based on numerical algorithms which optimize the layer thicknesses to fit a given target reflectance curve \cite{Binda:2004, Aquila:2006, Pelizzo:2009, Jiang:2013, Kozhevnikov:2015, Kuang:2018}.  However, for the same target curve, several different solutions are possible. This is because the algorithm outcome depends on many factors, including the initial conditions and the details of the optimization algorithm.  Despite these limitations, the excellent results obtained so far suggest that aperiodic MLs are very promising structures for future solar missions.
 
\section{A case study of innovative diagnostic of hot plasma}
The SXR flare lines were observed for the first time
 in 1969 by  OSO-5 \cite{kastner_etal:74},
 and provide excellent density diagnostics around  10$^{11}$  cm$^{-3}$.
 %, unavailable at any other spectral range. For example, 
 %the X-ray lines  are sensitive to much higher densities,
 %around 10$^{13}$  cm$^{-3}$, see \cite{delzanna_mason:2018}.
Such SXR lines were  recently reviewed using full-Sun spectra
of X-class flares from the Extreme Ultraviolet Variability Experiment (EVE) \cite{woods2012} on board \textit{SDO} \cite{delzanna_woods:2013}.
Table~\ref{tab:list_soft_xrays} lists the strongest transitions, and those
very sensitive to densities (dd in the Table) that are available between 90 \r{A} and 150 \AA. These lines are normally
very weak, but for densities around 10$^{11}$  cm$^{-3}$ or more have
intensities that become measurable, being a fraction of the strongest lines. 
The most reliable measurements of electron densities are obtained by measuring
the ratio of one of these lines with a strong line from the same ion.
\begin{table}[!htbp]
\caption{List of the strongest soft X-ray Fe flare lines, and a few of the main 
 density-diagnostic lines,  highlighted with a 'dd' in the last column.
$\lambda$ (\AA) is the  experimental  wavelength, 
T$_{\rm max}$ (in logarithm, K)  the approximate temperature of formation
of the ion in equilibrium. C1 and C2 indicate the lines selected for the two ML channels. bl indicates blended lines}
\begin{center} 
\begin{tabular}{@{}lllllllll@{}}
\hline
 Ion &  $\lambda$  & Transition & T$_{\rm max}$ & Notes & ML candidates\\ 
\hline
%\ion{Fe}{xx} &  93.781 & 2s$^2$ 2p$^3$ $^2$D$_{5/2}$ - 2s 2p$^4$ $^2$P$_{3/2}$ &  7.0 & \\
\ion{Fe}{xviii} &  93.923 & 2s$^2$ 2p$^5$ $^2$P$_{3/2}$ - 2s 2p$^6$ $^2$S$_{1/2}$ &  6.9 &  & {Y/Mo, Y/Pd}\\ 

% \ion{Fe}{xxi} & 97.864 & 2s$^2$ 2p$^2$ $^3$P$_{1}$ - 2s 2p$^3$ $^3$S$_{1}$ &  7.1 & \\ 
% \ion{Fe}{xvii} &  98.25 & 2s$^2$2p$^5$3s $^3$P$_{1}$--2p$^6$3s $^1$S$_{0}$ &  6.9 & \\
%\ion{Fe}{xxii} & 100.775 & 2s$^2$ 2p $^2$P$_{1/2}$ - 2s 2p$^2$ $^2$P$_{3/2}$ &  7.1 &   \\ %

\ion{Fe}{xix} &  101.55 & 2s$^2$ 2p$^4$ $^3$P$_{2}$ - 2s 2p$^5$ $^3$P$_{1}$ &  7.0 & \\
\ion{Fe}{xxi} &  102.217 & 2s$^2$ 2p$^2$ $^3$P$_{2}$ - 2s 2p$^3$ $^3$S$_{1}$ &  7.1 & dd \\

\ion{Fe}{xviii} & 103.948 & 2s$^2$ 2p$^5$ $^2$P$_{1/2}$ - 2s 2p$^6$ $^2$S$_{1/2}$ &  6.9 & C1 \\ 

%\ion{Fe}{xix}  & 106.317 &  2s$^2$ 2p$^4$ $^3$P$_{1}$ - 2s 2p$^5$ $^3$P$_{0}$ &  7.0 & weak \\

\ion{Fe}{xxi} &  108.118 & 2s$^2$ 2p$^2$ $^3$P$_{0}$ - 2s 2p$^3$ $^3$P$_{1}$ &  7.1 & weak \\

\ion{Fe}{xix} &  108.355 & 2s$^2$ 2p$^4$ $^3$P$_{2}$ - 2s 2p$^5$ $^3$P$_{2}$ &  7.0 &  C1 \\ 

%\ion{Fe}{xix} &  109.952 & 2s$^2$ 2p$^4$ $^3$P$_{0}$ - 2s 2p$^5$ $^3$P$_{1}$ &  7.0 & weak \\ % 2-7  
%\ion{Fe}{xx} &   110.627 & 2s$^2$ 2p$^3$ $^2$D$_{3/2}$ - 2s 2p$^4$ $^2$D$_{3/2}$ &  7.0 &  \\ % 2-9 

%\ion{Fe}{xix} &  111.695 & 2s$^2$ 2p$^4$ $^3$P$_{1}$ - 2s 2p$^5$ $^3$P$_{1}$ &  7.0 & weak \\ % 3-7 
%\ion{Fe}{xx} &   113.349 &  2s$^2$ 2p$^3$ $^2$D$_{5/2}$  2s 2p$^4$ $^2$D$_{5/2}$ &  7.0 &  \\ 

\ion{Fe}{xxii} &  114.410 & 2s$^2$ 2p $^2$P$_{3/2}$ - 2s 2p$^2$ $^2$P$_{3/2}$ &  7.1 & dd \\

%\ion{Fe}{xxii} &  116.268 & 2s$^2$ 2p $^2$P$_{3/2}$ - 2s 2p$^2$ $^2$S$_{1/2}$ &  7.1 & dd \\ 

%\ion{Fe}{xxii} &  117.154 & 2s$^2$ 2p $^2$P$_{1/2}$ - 2s 2p$^2$ $^2$P$_{1/2}$ &  7.1 &   \\
 % 1-8  
%\ion{Fe}{xxi} &  117.50 & 2s$^2$ 2p$^2$ $^3$P$_{1}$ - 2s 2p$^3$ $^3$P$_{1}$ &  7.1 & dd \\ 

%\ion{Fe}{xx} & 118.680  & 2s$^2$ 2p$^3$ $^4$S$_{3/2}$ - 2s 2p$^4$ $^4$P$_{1/2}$ &  7.0 &  \\ 

%\ion{Fe}{xix} &  119.983 & 2s$^2$ 2p$^4$ $^3$P$_{1}$ - 2s 2p$^5$ $^3$P$_{2}$ &  7.0 &  dd \\  3-6 

\ion{Fe}{xxi} &  121.213 & 2s$^2$ 2p$^2$ $^3$P$_{2}$ - 2s 2p$^3$ $^3$P$_{2}$ &  7.1 & C1 dd  \\ % 3-12 

\ion{Fe}{xx} &  121.845 & 2s$^2$ 2p$^3$ $^4$S$_{3/2}$ - 2s 2p$^4$ $^4$P$_{3/2}$ &  7.0 & C1 \\ 

%\ion{Fe}{xxi} &  123.831 & 2s$^2$ 2p$^2$ $^3$P$_{2}$ - 2s 2p$^3$ $^3$P$_{1}$ &  7.1 &  dd weak  \\ 

\hline

\ion{Fe}{xxi} &  128.753 & 2s$^2$ 2p$^2$ $^3$P$_{0}$ - 2s 2p$^3$ $^3$D$_{1}$ &  7.1 &  C2 &{Mo/Si, Mo/Be}\\ 

\ion{Fe}{xx} & 132.840 & 2s$^2$ 2p$^3$ $^4$S$_{3/2}$ - 2s 2p$^4$ $^4$P$_{5/2}$ &  7.0 & C2 bl \\ % 1-6 strongest
\ion{Fe}{xxiii} &  132.906 & 2s$^2$ $^1$S$_{0}$ - 2s 2p $^1$P$_{1}$ &  7.2 & C2 bl \\ % 3.57e+06

 \ion{Fe}{xxii} &  135.791 & 2s$^2$ 2p $^2$P$_{1/2}$ - 2s 2p$^2$ $^2$D$_{3/2}$ &  7.1 & C2 \\ 
 \ion{Fe}{xxi} &  142.144 & 2s$^2$ 2p$^2$ $^3$P$_{1}$ - 2s 2p$^3$ $^3$D$_{2}$ &  7.1 & C2 dd \\
\ion{Fe}{xxi} &  142.281 & 2s$^2$ 2p$^2$ $^3$P$_{1}$ - 2s 2p$^3$ $^3$D$_{1}$ &  7.1 & C2 dd   \\ 
 \ion{Fe}{xxi}  &  145.732  & 2s$^2$ 2p$^2$ $^3$P$_{2}$ - 2s 2p$^3$ $^3$D$_{3}$ &  7.1 & C2 dd \\
 \ion{Fe}{xxii}  &  156.019 &   2s$^2$ 2p $^2$P$_{3/2}$ - 2s 2p$^2$ $^2$D$_{5/2}$ &  7.1 & dd  \\
 \hline
\end{tabular}
\end{center}
\label{tab:list_soft_xrays}
\end{table}
% Table~\ref{tab:list_soft_xrays}

There are several choices of wavelengths, but to maximize the science
capabilities, an instrument should be able to simultaneously measure the strong lines
from each ionization stage, plus at least one or two density diagnostics. In order to prove the feasibility of such an instrument, a list of ten SXR lines has been selected as example of innovative hot plasma diagnostic. For this case study, the performance achievable with multiband MLs based on both M-fold and aperiodic structures has been investigated considering two separate spectroscopic channels, one having a coating with high reflectance in emission lines below 124 \r{A} (i.e. Si L$_{2,3}$-edge) and the other equipped with a coating with high efficiency between 124 \r{A} and 150 \r{A}. Table \ref{tab:list_soft_xrays} summarizes the lines selected for this work, specifying if they fall in the first or second channel (marked in the table with C1 or C2, respectively); the table also reports the ML material couples that are suitable for each channel. Furthermore, it has been assumed that the two channels share all the optical components in order to reduce the mass and the instrument dimension. The coating of each spectroscopic channel will thus take up half of each optical component, following the strategy successfully exploited, for example, in the spectrometer Hinode/EIS or the imaging telescopes of SDO/AIA.

The optimization of the ML structures has been carried out by using a handmade algorithm which was developed by adapting the algorithm proposed by \cite{Binda:2004}. Our algorithm identifies the best-performing ML among a population of solutions -- 500 in this study -- that evolves through random variation of a sub-set of the structure-identifying parameters. The optimization starts from a seeding structure and is stopped after a user-selected number of iterations -- 1000 in this study -- or if the merit function (MF) assumes values which are lower than a certain threshold. Here, the MF has been defined as:
\begin{equation}
\rm{MF}=\sum_{i=1}^{k}w_i\left[ R_{cur}(\lambda_i) - R_{targ}(\lambda_i)\right ]^2    
\end{equation}
where $R_{cur} (\lambda_i )$ is the current structure reflectance at $\lambda_i$, $ R_{targ}(\lambda_i)$ is the target (i.e. wanted) reflectance at $\lambda_i$ and $w_i$ is a weighting factor. For the M-fold stacks, the optimization consists in determining the thickness of the period $d$, the $\Gamma$ ratio, the number $N$ of periods for each section, as well as the thickness of the layer. These quantities represent the structure-identifying parameters to be determined by the optimization algorithm. In the case of the aperiodic stacks, the  thickness of the layers are the structure-identifying parameters. As an additional constraint, the solutions must have layers thicker than 1.5 nm. 

The coating optimization started by taking as seeding structure a three-fold ML whose initial values of period thickness and numbers and $\Gamma$ ratios were selected by independently tuning a periodic section for each target wavelength. Only cases with one or two matching layers were considered. After the optimization of the three-fold stack, the best performing solution in turn was used as a seeding structure for the aperiodic ML optimization. After the ML optimization, the normalized effective area $\overline{A_{eff}}$ of both channels of our hypothetical instrument (for which we assume a optical scheme similar to that of Hinode/EIS) was also estimated. Under the hypothesis of a good correction of the aberrations, the effective area $A_{eff}$ is given by
\begin{equation}
A_{eff}(\lambda)=\frac{\pi}{2} \left( \frac{\Phi}{2} \right)^2 T_{F}(\lambda) R_{P}(\lambda) R_{G}(\lambda) \eta(\lambda) 
\end{equation}
where $\Phi$ is the diameter of primary mirror, $T_{F}(\lambda)$ is the transmission of the entrance metallic filter, $R_{P}(\lambda)$ is the reflectance of the ML on the primary mirror, $R_G(\lambda)$ is the grating efficiency, $\eta(\lambda)$ is the quantum efficiency of the detector and the additional factor $\frac{1}{2}$ takes into account that each channel  exploits only half of the primary mirror area. The normalized effective area $\overline{A_{eff}(\lambda)}$ is then defined as
\begin{equation}
\overline{A_{eff}(\lambda)}=\frac{A_{eff}(\lambda)}{\Phi^2\eta(\lambda)}=\frac{\pi}{8}  T_{F}(\lambda) R_{P}(\lambda) R_{G}(\lambda) 
\end{equation}
which is a general parameter useful to summarize the obtainable performances without depending on primary mirror area or detector quantum efficiency. 

\begin{figure} [ht]
   \begin{center}
   \begin{tabular}{c} 
   \includegraphics[width=0.96\textwidth]{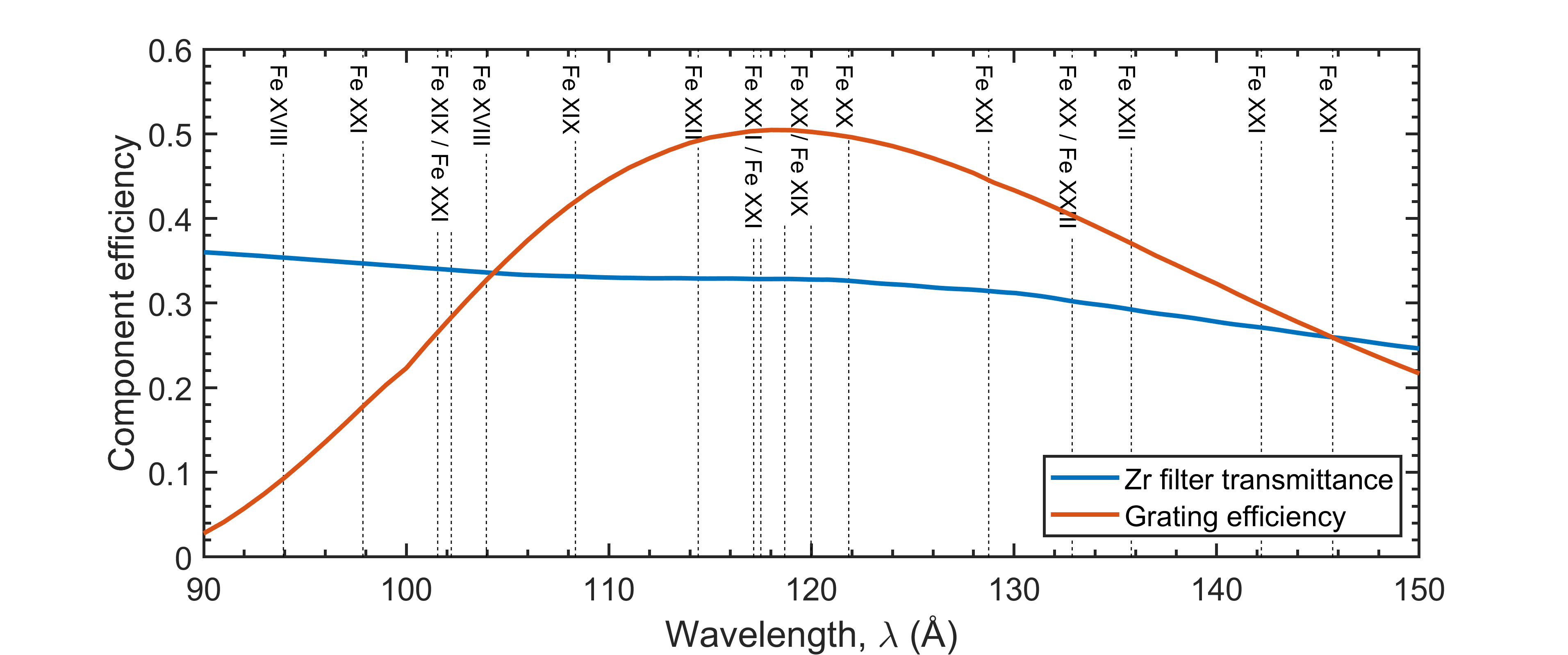}
	\end{tabular}
	\end{center}
   \caption[Transmittance of the entrance Zr filter (blue solid line) and grating efficiency considered for the normalized effective area estimation.] 
   {\label{fig:Zr_and_grooves}Simulated transmittance of the entrance Zr filter (blue solid line) and grating efficiency (orange solid line) that were used to estimate the normalized effective area.}
\end{figure}

The entrance filter is assumed to be a 240 nm thick film of Zr mounted on a supporting mesh. 
Zr filters have been used e.g. for the SXR telescopes of \textit{SDO}/AIA and have shown very little degradation, unlike most commonly-used filters in the EUV range \cite{Boerner:2014}.
In the simulation of the filter transmittance (see Figure \ref{fig:Zr_and_grooves}) it was taken into account an oxidation of 10 nm for both sides and an efficiency loss of 15\% for the supporting mesh (i.e. transmission of the mesh of 85\%). These parameters are taken from the values measured for the\textit{SDO}/AIA  94 and 131 \r{A} filters \cite{Boerner:2011}.
The grating efficiency was estimated by multiplying the ML reflectance by the groove efficiency computed with the PCGrate code. The choice of the grating structure depends on many aspects (e.g. optical scheme, expected field of view, instrument resolution, etc...), and therefore a general and unique solution cannot be given in the present work. However, a reasonable and representative solution based on the available literature can be provided. Following the work of \cite{Voronov:2010}, a grating of 5000 grooves/mm working with an incidence angle of 10$^\circ$ and a blaze angle of 5$^\circ$ has been selected as a representative case. Moreover, a non-ideal shape profile like that reported by Voronov was used in the grating simulations in order to improve the realism of the results. With this grating configuration, the relative incidence angle viewed from the ML is then 5$^\circ$ -- the MLs design angle considered in the present study -- and the blaze condition is satisfied for the third order. The grating efficiency used for the effective area estimation is reported in Figure \ref{fig:Zr_and_grooves}.

\begin{figure} [!ht]
   \begin{center}
   
   \includegraphics[width=0.96\textwidth]{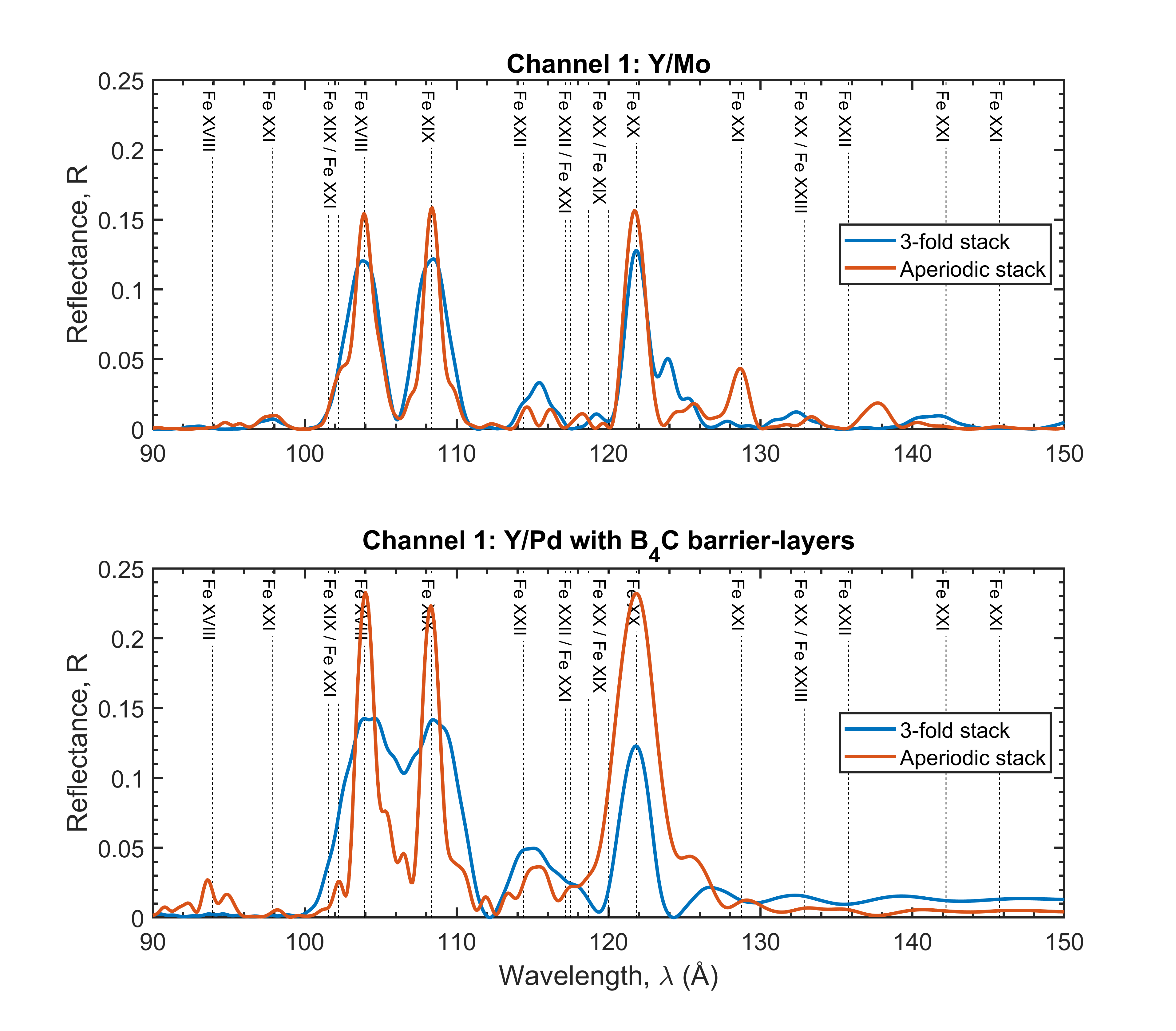}

	\end{center}
   \caption[Three-fold and aperiodic ML structures developed for the Channel 1.] 
   { \label{fig:Channel1} Three-fold (blue solid lines) and aperiodic (orange solid lines) ML stacks developed for the Channel 1 of the considered case study (see Table \ref{tab:list_soft_xrays}). The top panel shows the solution based on the standard and well known Mo/Y ML coating, whereas  the bottom panel shows the solution based on the innovative and promising Y/Pd ML with B$_4$C inter-layers. The simulations were performed using the optical constants available in the CXRO database \cite{CXRO:web}. The effects of the interfaces roughness and inter-diffusion were modelled using a Gaussian error function according to the N\'evot and Croce formalism \cite{Nevot:1980}. The Y/Mo interfaces were modelled with an error function with $\sigma =0.57$ nm \cite{Corso:2015} while the Pd/Y interfaces were modelled with an error functions with $\sigma =0.65$ nm and B$_4$C barrier-layers with a thickness of 0.6 nm \cite{Windt:2015}.}
\end{figure}

Figure \ref{fig:Channel1} reports the most promising results obtained for Channel 1 after the optimization process. 
In this channel, the Y/Pd coating with 0.6 nm-thick B$_4$C barrier-layers shows the best performance. Considering the 3-fold design, Y/Pd has a reflectance of 0.14 for the \ion{Fe}{xviii} (103.9 \AA) and \ion{Fe}{xix} (108.3 \AA) lines while the Mo/Y gives  a reflectance of 0.12. At the \ion{Fe}{xx} (121.8 \AA) line, Mo/Y is instead more efficient, having a reflectance of 0.13 against 0.12 of Y/Pd. Nevertheless, the aperiodic refinement considerably improves the Y/Pd performance, allowing to reach reflectance values of about 0.22 for all the three lines, as compared to a value of 0.15 which is reached by Y/Mo. Hence, in this channel the most promising strategy for increasing the final effective area is to use an aperiodic Y/Pd ML stack.

\begin{figure} [!ht]
   \begin{center}
   \begin{tabular}{c} 
   \includegraphics[width=0.96\textwidth]{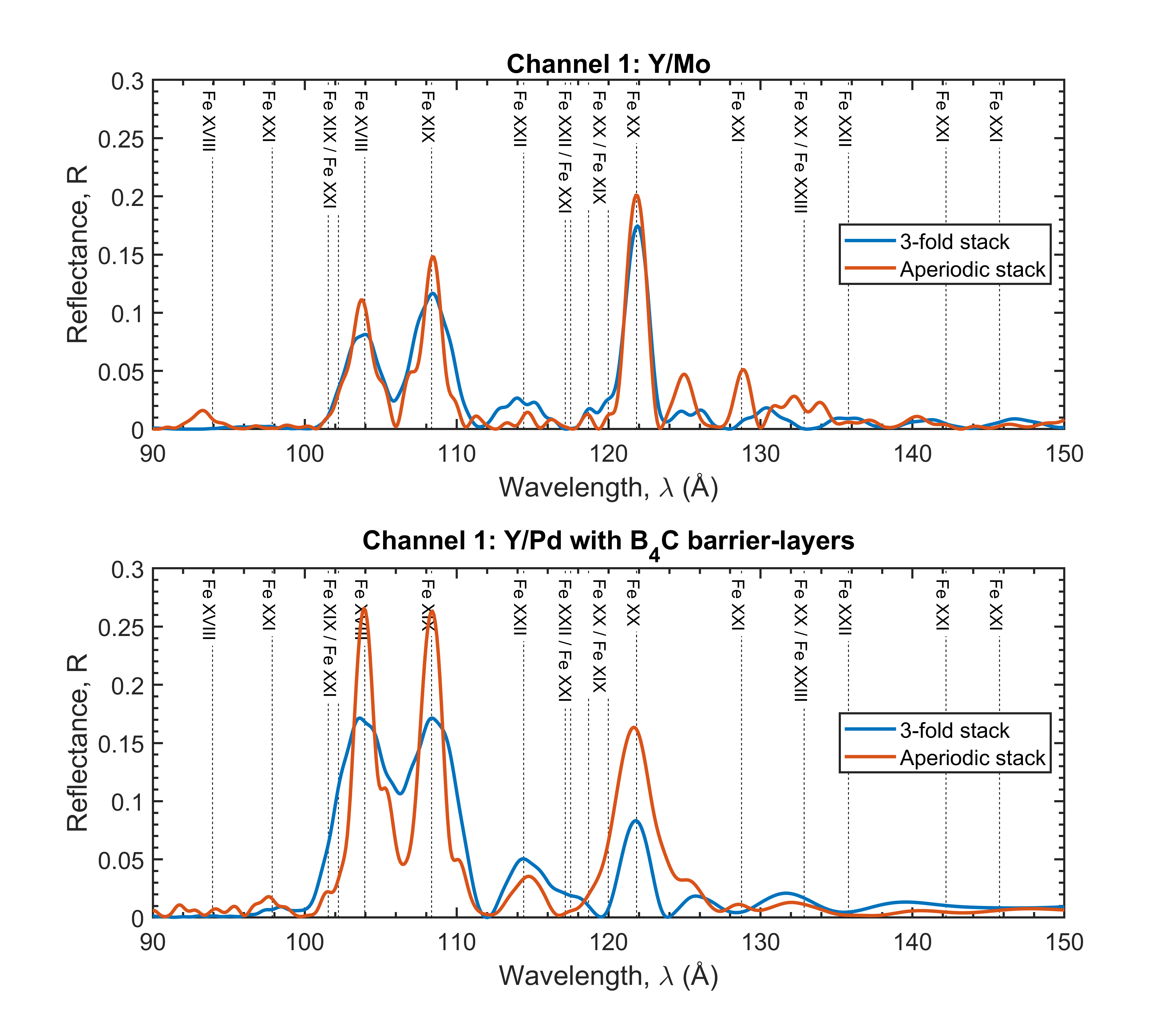}
	\end{tabular}
	\end{center}
   \caption[Alternative three-fold and aperiodic ML structures developed for the Channel 1.] 
   { \label{fig:ML_ch1_var} Alternative three-fold (blue solid lines) and aperiodic (orange solid lines) ML stacks developed for the Channel 1 of our case study (see Table \ref{tab:list_soft_xrays}). The solutions presented here optimize the reflectance peaks by exploiting the native efficiency of the material couples.}
\end{figure}

The maximum reflectance peaks achieved in a multiband ML depend on the native efficiency of the materials couple reported in Figure \ref{fig:MLsummary} as well as the number of required bands, their relative reflectance, and their mutual distance. Thus, once the set of wavelengths is defined, many solutions are available, depending on the choices and the compromises made during the instrument performance definition. The first aspect to consider is the maximum reflectance peak desired for each target wavelength. In the examples reported in Figure \ref{fig:Channel1}, we chose to have the same reflectance peak (within 1\%) for each selected line but other options are possible. For instance, since the Mo/Y structure offers a very good native efficiency above 110 \AA, it is possible to increase the peak reflectance at the \ion{Fe}{xx} line by sacrificing the ones at shorter wavelengths (i.e. especially \ion{Fe}{xviii}) where this materials couple has a lower native efficiency. This case is illustrated in the top panel of Figure \ref{fig:ML_ch1_var}. The three-fold structure can achieve a reflectance of 0.17 at \ion{Fe}{xx} but by assuming a reflectance of 0.08 for the \ion{Fe}{xviii}, the reflectance at \ion{Fe}{xix} does not change. With the aperiodic refinement the scenario improves slightly, as the \ion{Fe}{xx} reflectance goes up to 0.2 with 0.15 at \ion{Fe}{xix} and 0.11 at \ion{Fe}{xviii}.    
The same approach can be applied also to the Y/Pd ML coating, allowing the improvement of the reflectance peaks for all the lines having a wavelength shorter than 110 \r{A}, where the coating has the highest native efficiency. This is clearly possible by dropping the reflectance peak for the \ion{Fe}{xx}, as shown in the three-fold and aperiodic ML designs reported in bottom panel of Figure \ref{fig:ML_ch1_var}. If we accept a reflectance of 0.08 in the \ion{Fe}{xx} line, the three-fold stack can give a reflectance of 0.17 in the remaining two lines. With the aperiodic stack, the \ion{Fe}{xviii} and \ion{Fe}{xix} reflectance improves considerably since it achieves a value of 0.26 for both lines, although the reflectance at \ion{Fe}{xx} is limited to 0.16.

\begin{figure} [!ht]
   \begin{center}
   \begin{tabular}{c} 
   \includegraphics[width=0.96\textwidth]{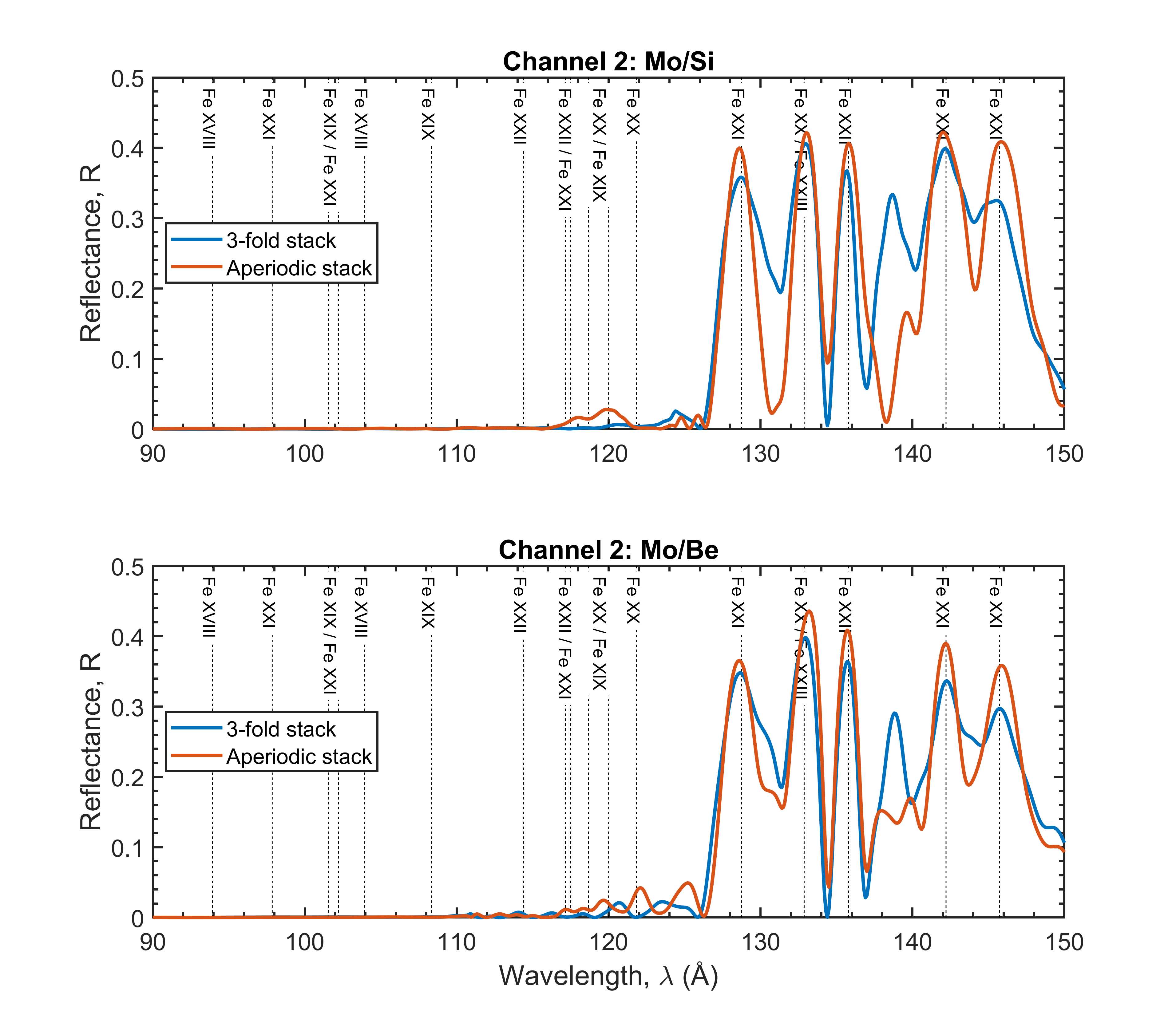}
	\end{tabular}
	\end{center}
   \caption[Three-fold and aperiodic ML structures developed for the Channel 2.] 
   { \label{fig:ML_ch2} Three-fold (blue solid lines) and aperiodic (orange solid lines) ML stacks developed for the Channel 2 of our case study (see Table \ref{tab:list_soft_xrays}). The top panel shows the solution based on the standard and well known Mo/Si ML coating, whereas the bottom panel shows the solution based on the innovative and promising Mo/Be ML. The simulations were performed using the optical constants available in the CXRO database \cite{CXRO:web}. The effects of the interfaces roughness and inter-diffusion were modelled using a Gaussian error function according to the N\'evot and Croce formalism \cite{Nevot:1980}. The Si/Mo interfaces were modelled with an error function with $\sigma =0.65$ nm \cite{Corso:2011}.For Mo/Be, two different error functions were used: one with a $\sigma =0.36$ nm for Be on Mo interfaces and $\sigma =0.7$ nm for Mo on Be interfaces \cite{Svechnikov:2018, Bajt:1999SPIE}.}
\end{figure}

In Channel 2, the two most promising material couples in term of native efficiency are the Si/Mo and Mo/Be. Figure \ref{fig:ML_ch2} reports the results obtained after the optimization process by using these two MLs. The coatings are characterized by a high reflectance in all target bands, mainly due to the high transparency offered by the spacer materials (i.e. Si or Be) which are both working close to their absorption edge. The Mo/Si seems to provide the best performance (Figure \ref{fig:ML_ch2}, top panel). The three-fold design gives a reflectance always greater than 0.32 for all five target spectral bands, with some peaks that even approach a value of 0.4: the \ion{Fe}{xxi} doublet at 142.1 \r{A}/142.3 \r{A} and \ion{Fe}{xx} at 132.8 \AA. The aperiodic refinement gives a fair improvement, which mainly brings the reflectance of all five bands to 0.4. The greatest improvement from the aperiodic design is obtained for the \ion{Fe}{xxi} line at 145.7 \AA, where the reflectance increases from 0.33 up to 0.4. This result perfectly agrees with the values reported by \cite{Gullikson:2015}, which described the development of an aperiodic Mo/Si coating for reflecting three different harmonics of a FEL beam in the 130 \r{A}--140 \r{A} wavelength range. 
The Mo/Be ML coating (Figure \ref{fig:ML_ch2}, bottom panel) provides reflectance curves very similar to those of Si/Mo. The three-fold design ensures a reflectance higher than 0.3 for all spectral bands, with the minimum value at 145.7 \AA. Also, in this case the aperiodic refinement improves slightly the performance, as the bands tuned at 132.4 \AA, 135.8 \r{A} and 142.3 \r{A} reach a reflectance of 0.4, whereas in the remaining two bands the ML shows a value of about 0.36. 
 
\begin{figure} [!ht]
   \begin{center}
   \begin{tabular}{c} 
   \includegraphics[width=0.96\textwidth]{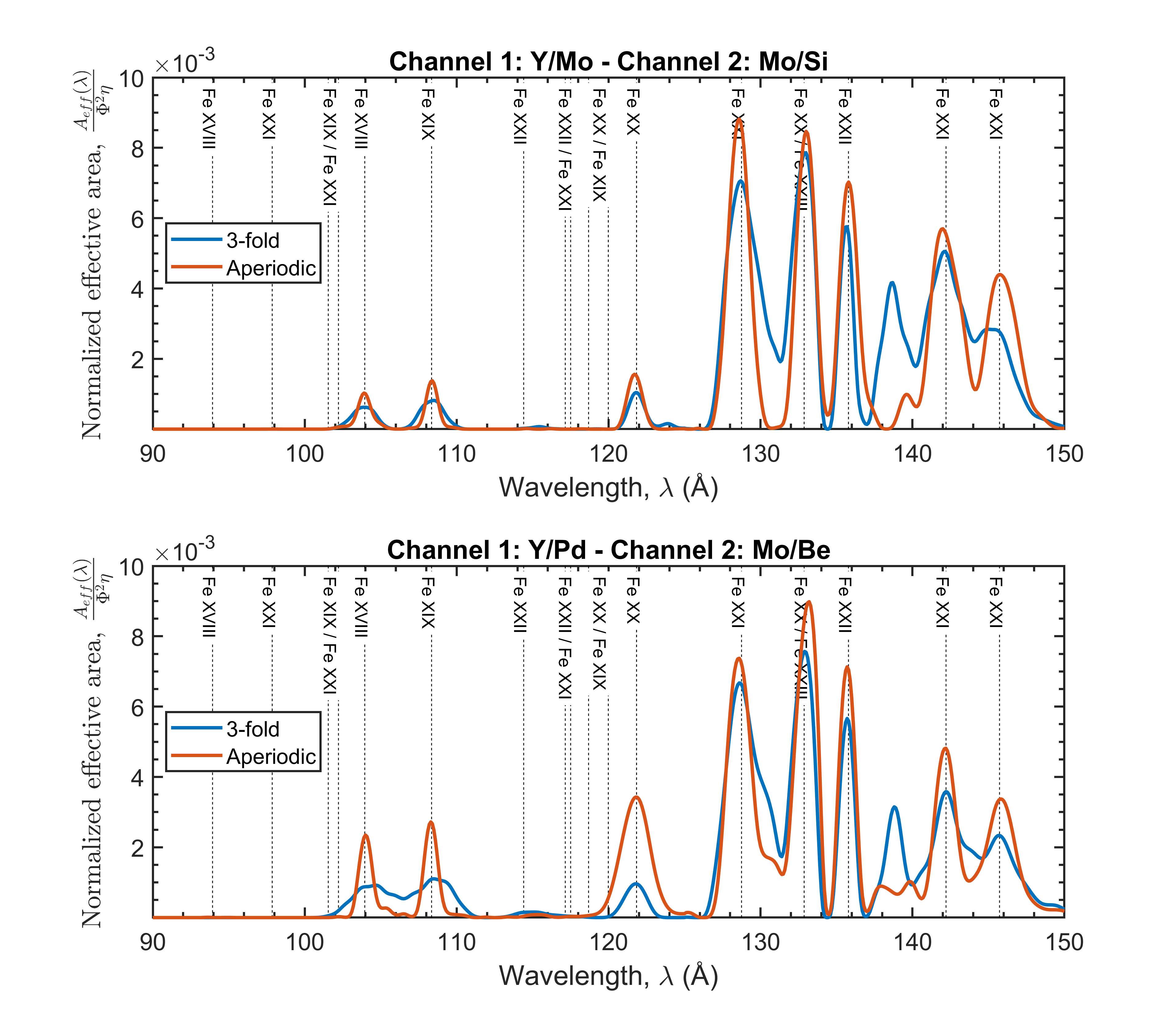}
	\end{tabular}
	\end{center}
   \caption[Normalized effective area $\bar{A_{eff}(\lambda)}$ obtainied employing three-fold or aperiodic ML structures developed for the case study considered in this paper.] 
   { \label{fig:EffArea} Normalized effective area $\overline{A_{eff}(\lambda)}$ obtainied employing three-fold (blue solid lines) or aperiodic (red solid lines) ML structures developed for the case study considered in this work. The normalized effective area presented in the top panel is based on the standard Mo/Y and Si/Mo MLs, whereas those in the bottom panel exploits the new Y/Pd and Mo/Be MLs.}
\end{figure}

Starting from the ML reflectance curves reported in Figure \ref{fig:Channel1} and Figure \ref{fig:ML_ch2}, an estimation of the normalized effective area $\overline{A_{eff}(\lambda)}$ has been performed considering both the three-fold and aperiodic designs (see Figure \ref{fig:EffArea}). In the top panel, the effective area curves have been computed considering the Mo/Y ML for the Channel 1 and the Si/Mo for the Channel 2; such ML materials are considered standard materials because they have already flown in many long term space missions. The bottom panel shows the effective area computed considering the new ML materials, Y/Pd with B$_4$C barrier-layers in Channel 1 and Mo/Be in Channel 2.
High values of normalized effective area are obtained by using a Y/Pd ML in Channel 1 and a Mo/Si ML in Channel 2. As expected, the aperiodic structures are the best performing but also the most challenging from a manufacturing point of view because the layers thickness layers are all different from each other. This suggests that it is important to evaluate the benefits that can be obtained from a certain structure, considering also the eventual drawbacks. For example, although the aperiodic design of the Si/Mo ML gives a substantial improvement for the lines at 128.7\r{A} and 145.7\r{A}, the normalized effective area reached by using the simpler three-fold structure is already very good for most scientific investigations. On the other hand, for Channel 1 the aperiodic Y/Pd ML is probably the best choice if one wishes to achieve a good effective area using mirrors with a small diameter. Therefore, the best solution in terms of both effective area and manufacturing complexity seems to be the use of a Y/Pd aperiodic stack with B$_4$C barrier-layers in Channel 1 and a Si/Mo three-fold stack in Channel 2.       
 
\section{Conclusions}
The SXRs are very promising for probing hot solar plasma, from about 3 MK up to 15MK. To answer many of the main scientific questions regarding the hot solar corona, it is crucial to obtain simultaneous spectroscopic observations at high resolution and high throughput in many different SXR emission lines. Spectrometers based on a normal incidence configuration are therefore highly recommended for this application as they are able to provide both high resolutions and valuable effective areas. Mirrors coated by nanometric MLs are certainly among the most important components in these instruments as they allow the development of high-performing normal-incidence optics working in this spectral range. So far, the vast majority of ML coatings employed in long-term missions have been based on periodic stacks, a type of structure that has shown great reliability, manufacturing repeatability, good temporal stability, and high reflectance but that unfortunately is characterised by a very narrow spectral band. However, in the last two decades, considerable improvements in the manufacturing of such MLs have been achieved, and today non-periodic stacks are reliable and  can provide high efficiency in broad or multi-band instrumentation. 

In this work, after giving an updated review of the materials and performance of the main non-periodic ML coatings available in the literature, the discussion has been focused on investigating the strategies that can be exploited to develop a SXR spectroscopic instrument capable of simultaneously observing eight different emission lines. The following set of lines was selected in order to maximize the scientific return: \ion{Fe}{xviii} (103.9 \AA), \ion{Fe}{xix} (108.3 \AA), \ion{Fe}{xx} (121.8 \AA) , \ion{Fe}{xxi} (128.7 \AA), \ion{Fe}{xxii} (135.8 \AA), and \ion{Fe}{xx}/\ion{Fe}{xxiii} (132.9 \AA) plus two different density-diagnostic lines (at 142.3 \r{A} and 145.7 \AA). The most promising configuration can be obtained by splitting the instrument into two channels, each of which exploits half of the optical components area. Channel 1 would be equipped with a multiband coating which allows to observe the three emission lines with  wavelength shorter than 124 \r{A} (i.e. Si L$_{2,3}$-edge); at these wavelengths, the most performing material couples are based on Mo/Y or Y/Pd with B$_4$C barrier-layers. Further, the coating of Channel 2 should be able to provide high efficiency at the remaining five lines. For this channel, the best material couples are the Mo/Si and Mo/Be. A generic algorithm was employed to optimize the ML reflectance for each channel. For each selected material couple, the best solutions based on both three-fold and aperiodic ML structures have been discussed, considering also different possible trade-offs. The three-fold version of the coatings is the simplest one from a manufacturing point of view, as this structure is obtained by the superposing three periodic blocks. The aperiodic version is slightly more challenging since the layers' thicknesses are different from each other. Moreover, a first order estimation of the  effective area, normalized with respect to the square of the primary mirror diameter and detector quantum efficiency, has been computed for a hypothetical instrument having a optical scheme similar to that of Hinode/EIS (i.e. an optical train composed by an entrance Zr filter, a ML-coated primary mirror, a slit and a ML-coated diffraction grating).

Our results show that the best performance is achieved by adopting a Y/Pd ML in Channel 1 and a Mo/Si ML in Channel 2; both these material couples have been deeply tested, showing high stability over time and thermal cycles. As expected, the higher reflectance curves (and normalized effective areas) are always achieved by aperiodic structures. However, comparing the Si/Mo three-fold structure with its aperiodic version clearly suggests that the performance improvement does not adequately justify the increased complexity of the aperiodic structure. In fact, the three-fold stack is able to give an overall normalized effective area that is already very good for most scientific investigations. In contrast, although the reflectance given by the three-fold Y/Pd stack can be suitable in many science cases, the significant improvement achievable with the aperiodic stack can be useful in many scientific contexts. Thus, the most reasonable solution in terms of effective area and manufacturing complexity seems to be the use of a Y/Pd aperiodic stack with B$_4$C barrier-layers in Channel 1 and a Si/Mo three-fold stack in Channel 2. Finally, giving the promising theoretical results presented in this work and the recent advancements in ML technology which have been achieved in the last few years, we suggest that the time is ripe for considering the development of SXR normal-incident spectroscopic instrumentation for solar physics applications.
  
%%% BIBLIOGRAPHY %%%%%%%%%%%%%%%%%%%%%%%%%%%%%%%%%%%%%%%%%%%%%%%%%%%%%%%%%%%

     % format of references provided by the journal (.bst)
%\bibliographystyle{spr-mp-sola}
     % name your Bibtex file containing your references (.bib)
\bibliographystyle{unsrt}

\bibliography{Main}  

     % Checking: look if the file containing the ``\bibitem'' exits
     %           so check if the .bbl file exist (bibTeX compilation)

\end{document}